\PassOptionsToPackage{dvipsnames,table,svgname,hyperref}{xcolor}

\documentclass[sigconf]{acmart}

\pdfoutput=1

\setcopyright{acmcopyright}
\copyrightyear{2018}
\acmYear{2018}
\acmDOI{XXXXXXX.XXXXXXX}

\acmConference[Conference acronym 'XX]{Make sure to enter the correct
  conference title from your rights confirmation emai}{June 03--05,
  2018}{Woodstock, NY}
\acmPrice{15.00}
\acmISBN{978-1-4503-XXXX-X/18/06}

\usepackage{multirow}
\usepackage{xspace}
\usepackage{soul}
\usepackage[ruled,vlined]{algorithm2e}
\SetAlFnt{\small}
\SetAlCapFnt{\small}
\SetAlCapNameFnt{\small}
\usepackage{amsmath}
\usepackage{bm}
\usepackage{hyperref}
\usepackage{soul}
\usepackage{tikz}
\usepackage{subcaption}
\usepackage{mathtools}
\usepackage{tabularx}

\usepackage{cleveref}
\usepackage{enumitem}
\usepackage{pgfplots}
\usepackage{makecell}
\usepackage{pgfplots}
\usepackage{subcaption}
\usepgfplotslibrary{colorbrewer}
\usetikzlibrary{calc,shadings,patterns}
\usetikzlibrary{fit}

\pgfplotsset{compat=1.18}

\newenvironment{customlegend}[1][]{%
\begingroup
\csname pgfplots@init@cleared@structures\endcsname
\pgfplotsset{#1}%
}{%
\csname pgfplots@createlegend\endcsname
\endgroup
}%
\def\addlegendimage{\csname pgfplots@addlegendimage\endcsname}

\usepackage{amsmath,scalerel}
\DeclareMathOperator*{\argmax}{arg\,max}

\DeclareMathOperator*{\concat}{\scalerel*{\Vert}{\sum}}

\definecolor{darkgray176}{RGB}{176,176,176}
\definecolor{darkorange25512714}{RGB}{255,127,14}
\definecolor{lightgray204}{RGB}{204,204,204}
\definecolor{steelblue31119180}{RGB}{31,119,180}

\author{Daniele Malitesta}
\authornote{Corresponding authors: Daniele Malitesta (\url{daniele.malitesta@poliba.it}) and Giandomenico Cornacchia (\url{giandomenico.cornacchia@poliba.it}).}
\affiliation{\institution{Politecnico di Bari, Italy}
  \city{}
  \country{}}
\email{daniele.malitesta@poliba.it}

\author{Giandomenico Cornacchia}
\authornotemark[1]
\affiliation{\institution{Politecnico di Bari, Italy}
  \city{}
  \country{}}
\email{giandomenico.cornacchia@poliba.it}

\author{Claudio Pomo}
\affiliation{\institution{Politecnico di Bari, Italy}
  \city{}
  \country{}}
\email{claudio.pomo@poliba.it}

\author{Tommaso {Di Noia}}
\affiliation{\institution{Politecnico di Bari, Italy}
  \city{}
  \country{}}
\email{tommaso.dinoia@poliba.it}

\begin{document}

\title{On Popularity Bias of Multimodal-aware Recommender Systems: a Modalities-driven Analysis}

\renewcommand{\shortauthors}{Daniele Malitesta, Giandomenico Cornacchia, Claudio Pomo, \& Tommaso Di Noia}






\begin{abstract}
Multimodal-aware recommender systems (MRSs) exploit multimodal content (e.g., product images or descriptions) as items' side information to improve recommendation accuracy. While most of such methods rely on factorization models (e.g., MFBPR) as base architecture, it has been shown that MFBPR may be affected by popularity bias, meaning that it inherently tends to boost the recommendation of popular (i.e., short-head) items at the detriment of niche (i.e., long-tail) items from the catalog. Motivated by this assumption, in this work, we provide one of the first analyses on how multimodality in recommendation could further amplify popularity bias. Concretely, we evaluate the performance of four state-of-the-art MRSs algorithms (i.e., VBPR, MMGCN, GRCN, LATTICE) on three datasets from Amazon by assessing, along with recommendation accuracy metrics, performance measures accounting for the diversity of recommended items and the portion of retrieved niche items. To better investigate this aspect, we decide to study the separate influence of each modality (i.e., visual and textual) on popularity bias in different evaluation dimensions. Results, which demonstrate how the single modality may augment the negative effect of popularity bias, shed light on the importance to provide a more rigorous analysis of the performance of such models.
\end{abstract}

\maketitle

\section{Introduction}\label{sec:introduction}

The massive availability of digital data (e.g., images, texts, audio tracks) on the Internet has recently favored the raising of a novel family of recommender systems (RSs), known as multimodal-aware recommender systems (MRSs). With the integration of multimodal features (extracted through pre-trained deep learning models~\cite{DBLP:conf/cvpr/HeZRS16, DBLP:conf/icassp/HersheyCEGJMPPS17, DBLP:conf/emnlp/ReimersG19}) as items' side information, MRSs can generate more accurate recommendations than traditional collaborative filtering~\cite{DBLP:conf/sigir/ChenCXZ0QZ19, DBLP:conf/mm/Zhang00WWW21, DBLP:conf/www/WeiHXZ23} (CF) algorithms by providing a countermeasure to common issues such as the sparsity of the user-item matrix and the cold-start scenario~\cite{DBLP:conf/aaai/HeM16,DBLP:conf/recsys/OramasNSS17,DBLP:conf/mm/VermaGGS20}, or the inexplicability of users' preferences in the implicit feedback setting~\cite{DBLP:conf/sigir/ChenZ0NLC17, DBLP:conf/mm/LiuCSWNK19, DBLP:conf/sigir/Li0YSCZS21, DBLP:conf/recsys/CornacchiaNR21,DBLP:conf/caise/CornacchiaDNPR21}. 

\input{figures/popularity}

The vast majority of MRSs are generally based upon the famous matrix factorization with bayesian personalized ranking (MFBPR) recommendation model. On the one hand, matrix factorization~\cite{DBLP:journals/computer/KorenBV09} (MF) is a latent-factor approach that maps users and items in the recommendation system to embeddings in the latent space and is trained to reconstruct the user-item interaction matrix via the dot product of the respective factors. On the other hand, bayesian personalized ranking~\cite{DBLP:conf/uai/RendleFGS09} (BPR) is an optimization schema that drives from the assumption that, for each user, the predicted score of positive (i.e., interacted) and negative (i.e., non-interacted) items should diverge. Given its simple implementation and efficacy, MFBPR has long constituted the backbone of recommendation algorithms in CF~\cite{DBLP:conf/www/HeLZNHC17, DBLP:conf/sigir/0001DWLZ020, DBLP:conf/cikm/MaoZWDDXH21}, not only for multimodal recommendation.

Nevertheless, recommender systems (such as MFBPR) may be affected by popularity bias~\cite{DBLP:journals/umuai/JannachLKJ15, DBLP:conf/recsys/AbdollahpouriBM17, DBLP:conf/recsys/Baeza-Yates20, DBLP:journals/ipm/BorattoFM21} (\Cref{fig:popularity}), as they tend to boost the recommendation of the items from the \textit{short-head} (i.e., the popular ones) at the expense of the items from the \textit{long-tail} (i.e., the niche ones). Tackling popularity bias in recommendation has primarily followed four directions~\cite{DBLP:journals/tois/0007D0F0023}: (i) regularization techniques~\cite{DBLP:conf/recsys/KamishimaAAS14, DBLP:conf/recsys/AbdollahpouriBM17, DBLP:conf/sigir/ChenXLYSD20}, (ii) adversarial learning~\cite{DBLP:conf/cikm/KrishnanSSS18}, (iii) causal graphs~\cite{DBLP:conf/kdd/WangF0WC21, DBLP:conf/sigir/ZhangF0WSL021, DBLP:conf/www/ZhengGLHLJ21}, and (iv) other item re-ranking approaches~\cite{DBLP:conf/aies/Abdollahpouri19, DBLP:conf/flairs/AbdollahpouriBM19}.

Despite the growing interest in popularity bias~\cite{DBLP:conf/ecir/AnelliDNMPP23,DBLP:journals/ipm/CornacchiaABNPR23} and potential solutions to address it, to date, very limited effort has been put into investigating \textbf{\textit{how multimodal side information in MRSs could amplify the negative effects of popularity bias}}. To the best of our knowledge, three recent works discussed the concept of bias in multimodal-aware recommendation. First,~\citet{DBLP:conf/mm/LiuTSYH22} take into account the bias towards a single modality in multimodal recommendation, and propose a solution based upon causal inference and counterfactual reasoning; however, the definition they provide about bias is conceptually different from the one of popularity bias. Then,~\citet{DBLP:conf/bias/KowaldL22} consider popularity bias in the case of multimedia recommendation datasets (e.g., MovieLens); however, they do not support their findings by testing recommender systems leveraging multimodal features as items' side information. Last,~\citet{DBLP:conf/kdd/MalitestaCPDN23} investigate how novelty and diversity metrics are influenced in multimodal recommendation, but without a finer-grained analysis on the impact of each single modality.

Driven from the assumptions above, and differently from the related literature, we propose one of the first analyses on how multimodal-aware recommender systems may amplify popularity bias in the produced recommendation lists. To this aim, we select four established and recent multimodal-aware recommender systems from the literature (i.e., VBPR~\cite{DBLP:conf/aaai/HeM16}, MMGCN~\cite{DBLP:conf/mm/WeiWN0HC19}, GRCN~\cite{DBLP:conf/mm/WeiWN0C20}, and LATTICE~\cite{DBLP:conf/mm/Zhang00WWW21}) and train them on three categories of the Amazon recommendation dataset~\cite{DBLP:conf/sigir/McAuleyTSH15} (i.e., \textit{Office}, \textit{Toys}, and \textit{Clothing}). Then, we evaluate the performance of the models by assessing metrics accounting for recommendation accuracy and popularity bias (the latter is measured through the diversity of recommendation lists and the percentage of retrieved items from the long-tail). Finally, to tailor our investigation, we focus on the separate impact of each multimodal side information (i.e., visual or textual) on popularity bias. To conduct this further study, we train the selected recommender systems when integrating either the visual or the textual modality as items' side information, and study the performance on single metrics and across pairs of metrics. 

We seek to answer: \textbf{RQ1.} How do multimodal-aware recommendation models behave in terms of accuracy, diversity, and popularity bias? \textbf{RQ2.} What is the influence of each modality (i.e., visual, textual, multimodal) on such performance measures? Results widely show that the integration of a single modality (with respect to the multimodal setting) is capable of amplifying the negative effects of popularity bias, paving the way to additional, more formal investigations on multimodal recommendation. We release the code at:~\url{https://github.com/sisinflab/MultiMod-Popularity-Bias}.
\section{Related Work}
This section outlines the related literature about multimodal learning and popularity bias in recommendation. First, we provide an overview of the most popular and recent advances in multimodal-aware recommendation, from which we select four representative approaches to analyze. Then, we summarize the concept of popularity bias, underlining how our work provides one of the first comprehensive investigations on popularity bias in multimodal recommendation at the granularity of modalities.

\noindent \textbf{Multimodal-aware recommendation.}
In various domains such as fashion~\cite{DBLP:conf/kdd/ChenHXGGSLPZZ19,DBLP:conf/sigir/ChenCXZ0QZ19,DBLP:conf/ecir/DeldjooNMM22}, music~\cite{DBLP:conf/sigir/ChengSH16,DBLP:conf/recsys/OramasNSS17,DBLP:conf/bigmm/VaswaniAA21}, food~\cite{DBLP:journals/tmm/MinJJ20,DBLP:journals/eswa/LeiHZSZ21,DBLP:journals/tomccap/WangDJJSN21}, and micro-video~\cite{DBLP:conf/mm/WeiWN0HC19,DBLP:journals/tmm/ChenLXZ21,DBLP:journals/tmm/CaiQFX22} recommendation, the multimodal content associated with items (e.g., product images and descriptions, or audio tracks) has demonstrated to greatly enhance the representational power of recommender systems. 

Following the latest advances in multimodal learning~\cite{DBLP:conf/icml/NgiamKKNLN11, DBLP:books/acm/18/BaltrusaitisAM18, DBLP:journals/pami/BaltrusaitisAM19}, multimodal-aware recommender systems (MRSs) aim to tackle some long-term open challenges in personalized recommendation such as data sparsity and cold-start~\cite{DBLP:conf/aaai/HeM16,DBLP:conf/recsys/OramasNSS17,DBLP:conf/mm/VermaGGS20}. Moreover, leveraging multimodal content can help reveal underlying user-item interactions and intents through attention mechanisms, contributing to the explainability of recommendations~\cite{DBLP:conf/sigir/ChenZ0NLC17,DBLP:conf/sigir/ChenCXZ0QZ19,DBLP:conf/mm/LiuCSWNK19,DBLP:journals/ipm/TaoWWHHC20,DBLP:conf/sigir/Li0YSCZS21}.

With the recent outbreak of graph neural networks in recommendation~\cite{DBLP:conf/sigir/0001DWLZ020, DBLP:conf/cikm/MaoZXLWH21, DBLP:conf/sigir/PengSM22}, several techniques have started integrating multimodality into the user-item bipartite graphs and knowledge graphs~\cite{DBLP:journals/tnn/ScarselliGTHM09,DBLP:journals/corr/abs-2104-13478,DBLP:conf/sigir/Wang0WFC19,DBLP:conf/sigir/0001DWLZ020, DBLP:journals/tmm/WangWYWSN23}, refining the multimodal representations of users and items through different approaches implementing the message-passing schema. While some early attempts involve simply injecting multimodal item features into the graph-based pipeline~\cite{DBLP:conf/kdd/YingHCEHL18}, more advanced techniques learn separate graph representations for each modality and disentangle users' preferences at the modality level~\cite{DBLP:journals/ipm/TaoWWHHC20,DBLP:conf/mm/WeiWN0C20,DBLP:conf/cikm/KimLSK22}. Recent approaches focus on uncovering multimodal structural differences among items in the catalog~\cite{DBLP:conf/mm/Zhang00WWW21,DBLP:conf/mm/LiuYLWTZSM21,DBLP:conf/mir/LiuMSO022}, in some cases by leveraging self-supervised~\cite{DBLP:conf/www/ZhouZLZMWYJ23, DBLP:conf/www/WeiHXZ23} and contrastive~\cite{DBLP:conf/sigir/Yi0OM22} learning.

In this work, we select four popular and recent approaches in multimodal recommendation, namely, VBPR~\cite{DBLP:conf/aaai/HeM16}, MMGCN~\cite{DBLP:conf/mm/WeiWN0HC19}, GRCN~\cite{DBLP:conf/mm/WeiWN0C20}, and LATTICE~\cite{DBLP:conf/mm/Zhang00WWW21}, and test their performance to assess the impact of (multi)modalities on popularity bias.

\noindent \textbf{Popularity bias in recommendation.}
In recommendation, popularity bias refers to the system's tendency to favor popular items (i.e., \textit{short-head}) at the expense of less popular ones (i.e., \textit{long-tail})~\cite{brynjolfsson2006niches, DBLP:journals/umuai/JannachLKJ15, DBLP:conf/recsys/AbdollahpouriBM17, DBLP:conf/recsys/Baeza-Yates20, DBLP:journals/ipm/BorattoFM21}. For instance,~\citet{DBLP:journals/umuai/JannachLKJ15} conduct a comprehensive algorithmic comparison across multiple datasets; their findings indicate that existing recommendation methods tend to concentrate mainly on a small fraction of the available item spectrum. More recently,~\citet{DBLP:conf/flairs/AbdollahpouriBM19} delve into this issue using the well-known MovieLens 1M dataset and reveal that over 80\% of all ratings are attributed to popular items; their main focus lies in finding ways to strike a balance between ranking accuracy and the coverage of long-tail items.

On such basis, the literature currently recognizes four main research directions~\cite{DBLP:journals/tois/0007D0F0023} to address popularity bias in recommendation, namely: (i) regularization techniques~\cite{DBLP:conf/recsys/KamishimaAAS14, DBLP:conf/recsys/AbdollahpouriBM17, DBLP:conf/sigir/ChenXLYSD20}, (ii) adversarial learning~\cite{DBLP:conf/cikm/KrishnanSSS18}, (iii) causal graphs~\cite{DBLP:conf/kdd/WangF0WC21, DBLP:conf/sigir/ZhangF0WSL021, DBLP:conf/www/ZhengGLHLJ21}, and (iv) other approaches such as item re-ranking~\cite{DBLP:conf/aies/Abdollahpouri19, DBLP:conf/flairs/AbdollahpouriBM19}. 

In multimodal recommendation, only a few recent works discuss popularity bias, but with specific definitions~\cite{DBLP:conf/mm/LiuTSYH22} and neglecting the impact of multimodal features~\cite{DBLP:conf/bias/KowaldL22}, or on other evaluation metrics~\cite{DBLP:conf/kdd/MalitestaCPDN23}. Conversely, our analysis assesses how prone multimodal-aware recommender systems are to push items belonging to the short-head and how the different modalities affect the tendency to amplify the popularity bias. 
\section{Background}
This section provides useful background notions for our proposed experimental analysis. To begin with, we introduce the preliminaries about the personalized recommendation scenario. Then, we focus on factorization-based approaches for recommendation (such as MFBPR) and present their building formulations. Finally, we extend the formalism to multimodal-aware recommendation, by considering the four selected approaches (i.e., VBPR, MMGCN, GRCN, and LATTICE) and their rationales. 

\subsection{Preliminaries}
Let $\mathcal{U}$ and $\mathcal{I}$ be the set of users and items in the recommendation system, respectively, where their cardinalities are indicated as $|\mathcal{U}|$ and $|\mathcal{I}|$. Then, let $\mathbf{X} \in \mathbb{R}^{|\mathcal{U}| \times |\mathcal{I}|}$ be the user-item interaction matrix, where $x_{ui} = 1$ if user $u$ interacted with item $i$, 0 otherwise. On such basis, we also introduce $\mathcal{R} = \{(u, i) \text{ }|\text{ } x_{ui} = 1\}$ as the set of recorded user-item interactions ($|\mathcal{R}|$ is its cardinality). 

\subsection{Factorization-based approaches} 
Currently, the majority of state-of-the-art recommender systems in collaborative filtering follow the matrix factorization~\cite{DBLP:journals/computer/KorenBV09} (MF) rationale. Despite the different building solutions they propose, the core idea is to map users' and items' IDs to embeddings in the latent space. Specifically, we indicate with $\mathbf{e}_u \in \mathbb{R}^d$ and $\mathbf{e}_i \in \mathbb{R}^d$ the embeddings for user $u$ and item $i$, respectively, with $d << |\mathcal{U}|, |\mathcal{I}|$. Then, given a pair of user and item $(u, i)$, the predicted interaction score is:
\begin{equation}
    \hat{x}_{ui} = \mathbf{e}_u^{\top} \mathbf{e}_i.
\end{equation}

To learn such embeddings, MF-based approaches are usually coupled with bayesian personalized ranking~\cite{DBLP:conf/uai/RendleFGS09} (BPR). This optimization method assumes that the predicted interaction score for users and their positive (i.e., interacted) items should be higher than the predicted interaction score for users and their negative (i.e., non-interacted) items. Concretely, let $\mathcal{T} = \{(u, i, j) \text{ } | \text{ } x_{ui} = 1 \land x_{uj} = 0\}$ be the set of triples, where each triple consists of a user, a positive, and a negative item. Bayesian personalized ranking seeks to optimize the following objective function:
\begin{equation}
    \argmax_{\Theta} \sum_{(u, i, j) \in \mathcal{T}} \text{ln } \sigma(\hat{x}_{ui} - \hat{x}_{uj}),
\end{equation}
where $\Theta$ is the vector containing all model's parameters (e.g., in the case of MF, $\mathbf{e}_u$ and $\mathbf{e}_i$), while $\sigma(\cdot)$ is the sigmoid function.

\subsection{Factorization-based approaches leveraging multimodal side information}
We present the formulations of four state-of-the-art multimodal-aware recommender systems (MRSs): VBPR~\cite{DBLP:conf/aaai/HeM16}, MMGCN~\cite{DBLP:conf/mm/WeiWN0HC19}, GRCN~\cite{DBLP:conf/mm/WeiWN0C20}, and LATTICE~\cite{DBLP:conf/mm/Zhang00WWW21}. Before diving into their approaches, we introduce some additional formalism. 

Besides $\mathbf{e}_u$ and $\mathbf{e}_i$, hereafter referred to as \textit{collaborative} user and item embeddings, we also introduce $\mathbf{f}_u$ and $\mathbf{f}_i$ as the \textit{multimodal} embeddings for user $u$ and item $i$. Moreover, we indicate $\mathcal{M}$ as the set of available modalities (e.g., visual, textual, audio), and we use $m$ as embedding's apex to denote that the embedding refers to the $m \in \mathcal{M}$ modality (e.g., $\mathbf{f}^m_i$ stands for the $m$-th multimodal embedding of item $i$).

\noindent\textbf{VBPR.}
Visual-bayesian personalized ranking~\cite{DBLP:conf/aaai/HeM16} (dubbed as VBPR) adopts visual features extracted from product images as items' side information in MFBPR. The authors introduce, along with user and item \textit{collaborative} embeddings, additional \textit{visual} user and item embeddings, where the latter is obtained as the activation of the penultimate layer from a pre-trained convolutional neural network. Then, the collaborative and visual embeddings are used to measure a collaborative- and visual-aware prediction for the interaction score and are eventually summed to obtain the final prediction score. In this work, we follow~\cite{DBLP:conf/mm/Zhang00WWW21} and adapt VBPR to multimodality by concatenating the visual and textual item features to generate a unique multimodal representation of the item:
\begin{equation}
    \hat{x}_{ui} = \mathbf{e}_u^{\top} \mathbf{e}_i + \mathbf{f}_u^{\top} t(\mathbf{f}_i) \quad \text{with} \quad \mathbf{f}_i = \concat_{m \in \mathcal{M}} \mathbf{f}^m_i,
\end{equation}
where $t$ is a projection function such that the latent dimensions of the multimodal user and item embeddings match.

\noindent\textbf{MMGCN.}
One of the first approaches leveraging the representational power of graph convolutional networks (GCNs) with multimodal content is multimodal graph convolution network for recommendation~\cite{DBLP:conf/mm/WeiWN0HC19} (dubbed as MMGCN). By designing one GCN for each modality, the model learns the different preferences users have towards each representation of the items. Finally, to fuse all multimodal representations into one for both users and items embeddings, the authors adopt the element-wise addition, and the predicted interaction score is calculated via the dot product:
\begin{equation}
    \hat{x}_{ui} = \mathbf{f}_u^{\top} \mathbf{f}_i \quad \text{with} \quad 
    \mathbf{f}_u = \sum_{m \in \mathcal{M}} c(\mathbf{e}_u, g(\mathbf{f}^{m}_u), t(\mathbf{f}^{m}_u, \mathbf{e}_u)),
\end{equation}
where $c$ and $g$ are a combination and GCN-based functions. We report only the user-side formulation for the sake of space.

\noindent\textbf{GRCN.}
Similarly to MMGCN, graph-refined convolutional network for multimedia recommendation~\cite{DBLP:conf/mm/WeiWN0C20} (dubbed as GRCN) utilizes a GCN-architecture to update user and item embeddings. Specifically, the adjacency matrix entries are refined by pruning the noisy user-item interactions according to the preference of users toward each item's modality. Collaborative and multimodal versions of the user and item embeddings are eventually combined through concatenation to estimate the interaction score via their dot product:
\begin{equation}
    \hat{x}_{ui} = \mathbf{f}_u^{\top} \mathbf{f}_i \quad \text{with} \quad \mathbf{f}_u = g(\mathbf{e}_u, \mathbf{f}^m_u, \forall m \in \mathcal{M})\; ||\; \left(\concat_{m \in \mathcal{M}} t(\mathbf{f}_u^m)\right).
\end{equation}
Again, we report only the user-wise formulation for lack of space.

\noindent\textbf{LATTICE.} Latent structure mining method for multimodal recommendation~\cite{DBLP:conf/mm/Zhang00WWW21} (dubbed as LATTICE) performs graph structure learning on multiple modality-aware item-item graphs (one for each modality). The obtained adjacency matrices are aggregated through weighted element-wise addition, and the final adjacency matrix is exploited to perform graph convolution to update the representation of the collaborative item embeddings. Then, this updated version is added to the initial collaborative item embedding. Finally, the dot product between the collaborative user and (updated) item embeddings predicts the interaction score:
\begin{equation}
    \hat{x}_{ui} = \mathbf{e}_u^{\top}\mathbf{f}_i \quad \text{with} \quad \mathbf{f}_i = \mathbf{e}_i + \frac{g(\mathbf{e}_i, \mathbf{f}^m_i, \forall m \in \mathcal{M})}{||g(\mathbf{e}_i, \mathbf{f}^m_i, \forall m \in \mathcal{M})||_2},
\end{equation}
where $g$ is a LightGCN~\cite{DBLP:conf/sigir/0001DWLZ020} architecture performing graph structure learning as stated above.

\section{Proposed analysis}

In this section, we present the details to conduct our analysis. Initially, we report on the used datasets, describing the methodologies employed for extracting multimodal features. Subsequently, we introduce and formally define the evaluation metrics employed, encompassing accuracy, diversity, and popularity bias. Finally, we provide a thorough summary of the reproducibility information for our study, detailing the methods used for dataset splitting and filtering as well as the strategy for hyperparameter search.

\subsection{Datasets}
The multimodal recommender systems have been tested on three popular~\cite{DBLP:conf/sigir/ChenCXZ0QZ19, DBLP:conf/mm/Zhang00WWW21, DBLP:conf/cikm/KimLSK22, DBLP:conf/www/ZhouZLZMWYJ23} datasets from the Amazon catalog~\cite{DBLP:conf/sigir/McAuleyTSH15}: Office Products (\textit{Office}), (b) Toys \& Games (\textit{Toys}), and (c) Clothing, Shoes \& Jewelry (\textit{Clothing}). The multimodal datasets provide both images and descriptions for each available item. Specifically, we utilize the pre-extracted 4,096-dimensional visual features~\cite{DBLP:conf/cvpr/DeldjooNMM21} which are made publicly available\footnote{\url{https://cseweb.ucsd.edu/~jmcauley/datasets/amazon/links.html}.}. For the textual modality, we follow the existing literature~\cite{DBLP:conf/mm/Zhang00WWW21}, which aggregates the item's title, descriptions, categories, and brand, thereby generating textual embeddings by leveraging sentence transformers~\cite{DBLP:conf/emnlp/ReimersG19}. The generated features are 1,024-dimensional embeddings. Additional dataset information can be found in~\Cref{tab:datasetInfo}.

\begin{table}[!t]
    \caption{Statistics of the tested datasets.}\label{tab:datasetInfo}
    \centering
    \begin{tabular}{lcccc}
    \toprule
        \textbf{Datasets} & $\bm{|\mathcal{U}|}$ & $\bm{|\mathcal{I}|}$ & $\bm{|\mathcal{R}|}$ & \textbf{Sparsity (\%)}\\ \cmidrule{1-5}
        \textit{Office} & 4,905 & 2,420 & 53,258 & 99.5513 \\
        \textit{Toys} & 19,412 & 11,924 & 167,597 & 99.9276 \\ 
        \textit{Clothing} & 39,387 & 23,033 & 278,677 & 99.9693 \\
        \bottomrule
    \end{tabular}
\end{table}

\subsection{Evaluation metrics}
In the proposed study, we refer to various metrics that may bring out additional insights which have not been investigated yet in multimodal recommendation. Indeed, we do not solely rely on accuracy metrics (i.e., Recall and nDCG) but also on diversity (i.e., item coverage) and popularity bias (i.e., APLT) metrics. The metrics listed hereinafter are calculated on top-$k$ recommendation lists.

\noindent\textbf{Recall.} The Recall assesses the system's capacity to retrieve relevant items from the recommendation list, highlighting the need for thorough coverage to the list of user interactions~\cite{DBLP:books/aw/Baeza-YatesR2011}: 
\begin{equation}\label{eq:Recall}
\mathrm{Recall} @ k=\frac{1}{|\mathcal{U}|} \sum_{u \in \mathcal{U}} \frac{\left|\operatorname{Rel}_u @ k\right|}{\left|\operatorname{Rel}_u\right|},
\end{equation}
where $\operatorname{Rel}_u$ indicates the set of relevant items for user $u$, while $\operatorname{Rel}_u @ k$ is the set of relevant recommended items in the top-$k$ list.

\noindent\textbf{Normalized discount cumulative gain.} The normalized discount cumulative gain (nDCG) considers the relevance and the ranking position of recommended products, taking into account the varied degrees of relevance:
\begin{equation}\label{eq:nDCG}
\mathrm{nDCG}@k=\frac{1}{|\mathcal{U}|} \sum_u \frac{{\mathrm{DCG}_u@k}}{\mathrm{IDCG}_u @ k}, 
\end{equation}
where $\mathrm{DCG}@k = \sum_{i=1}^{k} \frac{{2^{rel_{u,i}} - 1}}{{\log_2(i+1)}}$ quantifies the cumulative gain of relevance scores through the recommended list, with $rel_{u,i} \in \operatorname{Rel}_u$, and $\mathrm{IDCG}$ represents the cumulative gain of relevance scores for a perfect (ideal) recommender system.

\noindent\textbf{Item coverage.}
The item coverage (abbreviated ``iCov'' in the following) gives information on the coverage (item-side) measured in recommendation lists. A higher item coverage suggests that a larger fraction of the item space is being scrutinized and recommended to consumers, implying a more comprehensive coverage of user preferences and potentially a more comprehensive recommendation experience. In particular, we have:
\begin{equation}\label{eq:Coverage}
\mathrm{iCov}@k = \frac{{|\bigcup_{u} \hat{\mathcal{I}}_u@k|}}{{|\mathcal{I}_{train}|}},
\end{equation}
where $\hat{\mathcal{I}}_u@k$ is the list of top-$k$ recommended items for a user $u$.

\noindent\textbf{Average percentage of long-tail items.} \label{sec:aplt}
The average percentage of long-tail items (APLT) is a measure used to assess the presence of popularity bias in recommendation systems~\cite{DBLP:conf/recsys/AbdollahpouriBM17}. Popularity bias refers to the tendency of recommendation algorithms to prioritize popular or mainstream items over less well-known or niche items. This bias can lead to limited exposure of users to diverse and personalized recommendations. The metric measure the percentage of items belonging to the medium/long-tail distribution in the recommendation lists averaged over all users:
\begin{equation}
\mathrm{APLT}@k=\frac{1}{\left|\mathcal{U}\right|} \sum_{u \in \mathcal{U}} \frac{|\{i\; |\; i \in(\hat{\mathcal{I}}_u@k \;\cap\; \sim\Phi)\}|}{k},
\end{equation}
where $\Phi$ is the set of items belonging to the short-tail distribution while $\sim\Phi$ is the set of items from the medium/long-tail distribution. Note that we decide to integrate the evaluation of the APLT along with the iCov (introduced above) because the latter may be functional to provide a complete interpretation of the former. Indeed, following their definitions and formulations, the two metrics are conceptually related.

\noindent\textbf{Metrics value interpretation} 
An ideal recommender system should increase all the metrics listed above according to the principle ``higher is better'' to boost accuracy and diversity while reducing the popularity bias of the produced recommendations. \textit{\textbf{Nevertheless, with the current work, we try to unveil whether and why multimodal-aware recommender systems are affected by popularity bias. Thus, in the following, we will take into account those settings in which accuracy is high, while diversity and popularity bias are low (according to the metrics definitions).}}

\subsection{Reproducibility}
We investigate the models' behavior in three different settings: (i) \textit{visual} modality, in which we employ only visual features, (ii) \textit{textual} modality, in which we employ only textual features, and (iii) \textit{multimodal}, where both modalities are considered and combined.

In the first step, we evaluate the models in the multimodal setting which is the same setting as the original one for each tested approach. Then, we focused on quantifying the singular modality influence on the multimodal scenario in terms of accuracy, diversity, and popularity bias.
Furthermore, to ensure the reproducibility of our work, in the following, we provide comprehensive details regarding the preprocessing and splitting of the datasets, as well as the tuning and evaluation of the models.

The datasets are filtered using the $p$-core strategy, where we set $p$ to 5. Subsequently, we employ an 80\%/20\% train-test hold-out strategy to split the dataset. During the hyper-parameter tuning phase, we further divide the test set by removing 50\% of its instances for the validation, specifically evaluating the results using the Recall@20 metric (as in the original work). In terms of models' training, we set the maximum number of epochs to 200 and select the model weights based on the epoch that yields the best performance on the validation set.

The code is implemented in Elliot~\cite{DBLP:conf/sigir/AnelliBFMMPDN21}. Note that the explored hyper-parameter values are not entirely aligned with the ones in the original papers and codes. Indeed, we want to tune the selected baselines \textit{\textbf{on an extensive, shared set of hyper-parameter values across all models for the sake of fair comparison.}}
\section{Results and Discussion}
In this section, we answer the following research questions (RQs):

\begin{itemize}
    \item[\textbf{RQ1.}] \ul{\textit{How do the selected multimodal-aware recommendation models behave in terms of accuracy, diversity, and popularity bias?}} \Cref{sec:RQ1} investigates the recommendation performance in terms of accuracy (i.e., Recall, nDCG), diversity (i.e., iCov), and popularity bias (i.e., APLT). Note that, for the sake of completeness, we also report the performance of a recommender system generating recommendations in a random manner (i.e., Random) or based upon the most popular items in the catalog (i.e., MostPop); then, we train and evaluate MFBPR, that is the building model of the other multimodal baselines. We regard the performance of Random, MostPop, and MFBPR as a reference for the other multimodal-aware recommender systems we want to analyze.
    \item[\textbf{RQ2.}] \ul{\textit{What is the influence of each modality setting (i.e., visual, textual, multimodal) on such performance measures?}} \Cref{sec:RQ2} takes a step further by analyzing how each modality (i.e., visual, textual, and multimodal) influences accuracy, diversity, and popularity bias; the evaluation is conducted both on the single metric and across pairs of metrics.
\end{itemize}

\subsection{Recommendation accuracy, diversity, and popularity bias (RQ1)}\label{sec:RQ1}
\begin{table*}[!t]
\caption{Results in terms of recommendation accuracy (Recall, nDCG), diversity (iCov) and popularity bias (APLT). For accuracy metrics, $\uparrow$ means better performance, while $\downarrow$ means less diversity and more popularity bias. We remind that, while iCov and APLT metrics would generally adhere to the principle of ``higher is better'' ($\uparrow$) for an ideal recommender system, in this work we consider the opposite as we want to emphasize which models are performing worst in terms of diversity and popularity bias.}\label{tab:rq1}
\centering
\small
\begin{tabular}{llccrcccrcccrc}
\toprule
\multirow{2}{*}{\textbf{Datasets}} & \multirow{2}{*}{\textbf{Models}} & \multicolumn{4}{c}{top@10} & \multicolumn{4}{c}{top@20} & \multicolumn{4}{c}{top@50} \\ \cmidrule(lr){3-6} \cmidrule(lr){7-10} \cmidrule(lr){11-14} \multicolumn{2}{c}{} & \textbf{Recall}$\uparrow$ & \textbf{nDCG}$\uparrow$ & \textbf{iCov}$\downarrow$ & \textbf{APLT}$\downarrow$ & \textbf{Recall}$\uparrow$ & \textbf{nDCG}$\uparrow$ & \textbf{iCov}$\downarrow$ & \textbf{APLT}$\downarrow$ & \textbf{Recall}$\uparrow$ & \textbf{nDCG}$\uparrow$ & \textbf{iCov}$\downarrow$ & \textbf{APLT}$\downarrow$ \\ \cmidrule{1-14}
\multirow{8}{*}{\textit{Office}} & Random & 0.0034 & 0.0020 & 2,414 & 0.5950 & 0.0079 & 0.0034 & 2,414 & 0.5948 & 0.0220 & 0.0068 & 2,414 & 0.5924 \\
& MostPop & 0.0302 & 0.0208 & 20 & 0.0000 & 0.0533 & 0.0282 & 32 & 0.0000 & 0.1143 & 0.0439 & 66 & 0.0000 \\
& MFBPR & 0.0602 & 0.0389 & 2,268 & 0.2294 & 0.0955 & 0.0500 & 2,357 & 0.2379 & 0.1657 & 0.0677 & 2,398 & 0.2513 \\
\cmidrule{2-14}
& VBPR & \underline{0.0652} & \underline{0.0419} & 2,265 &  0.2321 & \underline{0.1025} & \underline{0.0533} & 2,354 &  0.2375 & \underline{0.1774} & \underline{0.0721} & 2,404 &
0.2469 \\
& MMGCN & 0.0455 & 0.0300 & \textbf{74} & \textbf{0.0016} & 0.0798 & 0.0405 & \textbf{112} & \textbf{0.0078} & 0.1575 & 0.0598 & \textbf{247} & \textbf{0.0205} \\
& GRCN & 0.0393 & 0.0253 & 2,390 & 0.3438 & 0.0667 & 0.0339 & 2,409 & 0.3469 & 0.1250 & 0.0488 & 2,414 & 0.3548 \\
& LATTICE & \textbf{0.0664} & \textbf{0.0449} & \underline{2,121} & \underline{0.1752} & \textbf{0.1029} & \textbf{0.0566} & \underline{2,315} & \underline{0.2039} & \textbf{0.1780} & \textbf{0.0751} & \underline{2,397} & \underline{0.2413} \\
\midrule
\multirow{8}{*}{\textit{Toys}} & Random & 0.0011 & 0.0006 & 11,879 & 0.4894 & 0.0021 & 0.0008 & 11,879 & 0.4896 & 0.0051 & 0.0015 & 11,879 & 0.4902 \\
& MostPop & 0.0130 & 0.0075 & 13 & 0.0000 & 0.0229 & 0.0104 & 24 & 0.0000 & 0.0451 & 0.0156 & 56 & 0.0000 \\
& MFBPR & 0.0641 & 0.0403 & 10,016 & 0.1167 & 0.0903 & 0.0481 & 10,944 & 0.1268 & 0.1394 & 0.0596 & 11,544 & 0.1460 \\
\cmidrule{2-14}
& VBPR & \underline{0.0710} & \underline{0.0458} & 10,085 &  0.1064 & \underline{0.1006} & \underline{0.0545} & 11,026 &  0.1180 & \underline{0.1523} & \underline{0.0667} & 11,624 & 0.1400 \\
& MMGCN & 0.0256 & 0.0150 & \textbf{4,499} & \underline{0.0961} & 0.0426 & 0.0200 & \textbf{6,238} & \underline{0.1058} & 0.0785 & 0.0285 & \textbf{8,657} & \underline{0.1263} \\
& GRCN & 0.0554 & 0.0354 & 11,007 & 0.2368 & 0.0831 & 0.0436 & 11,609 & 0.2482 & 0.1355 & 0.0559 & 11,847 & 0.2679 \\
& LATTICE & \textbf{0.0805} & \textbf{0.0512} & \underline{8,767} & \textbf{0.0546} & \textbf{0.1165} & \textbf{0.0617} & \underline{10,285} & \textbf{0.0684} & \textbf{0.1771} & \textbf{0.0759} & \underline{11,397} & \textbf{0.0950} \\
\midrule
\multirow{8}{*}{\textit{Clothing}} & Random & 0.0004 & 0.0002 & 23,016 & 0.4487 & 0.0010 & 0.0003 & 23,016 & 0.4478 & 0.0024 & 0.0006 & 23,016 & 0.4482 \\
& MostPop & 0.0089 & 0.0046 & 13 & 0.0000 & 0.0157 & 0.0063 & 24 & 0.0000 & 0.0322 & 0.0095 & 56 & 0.0000 \\
& MFBPR & 0.0303 & 0.0156 & 18,414 & 0.0729 & 0.0459 & 0.0195 & 20,582 & 0.0824 & 0.0734 & 0.0249 & 22,171 & 0.1017 \\
\cmidrule{2-14} 
& VBPR & \underline{0.0339} & \underline{0.0181}
& 19,195 & 0.0809 & \underline{0.0529} &
\underline{0.0229} & 21,251 & 0.0915 & 0.0847 & \underline{0.0292} & 22,555 & 0.1112
 \\
& MMGCN & 0.0227 & 0.0119 & \textbf{1,744} & \textbf{0.0044} & 0.0348 & 0.0150 & \textbf{2,864} & \textbf{0.0066} & 0.0609 & 0.0201 & \textbf{5,373} & \textbf{0.0121} \\
& GRCN & 0.0319 & 0.0164 & 21,490 & 0.2358 & 0.0496 & 0.0209 & 22,503 & 0.2459 & \underline{0.0858} & 0.0281 & 22,954 & 0.2631 \\
& LATTICE & \textbf{0.0502} & \textbf{0.0275} & \underline{13,463} & \underline{0.0134} & \textbf{0.0744} & \textbf{0.0336} & \underline{17,538} & \underline{0.0207} & \textbf{0.1186} & \textbf{0.0425} & \underline{21,458} & \underline{0.0385} \\
\bottomrule
\end{tabular}
\end{table*}

The results of the accuracy, diversity, and popularity bias metrics are reported in~\Cref{tab:rq1}. The measured values refer to top@10, top@20, and top@50 recommendation lists. In the following, we discuss the obtained results considering the three metrics families separately.

\noindent\textbf{Accuracy.} Overall, LATTICE is the top-performing model, in alignment with the recent literature~\cite{DBLP:conf/mm/Zhang00WWW21}. Indeed, its ability to learn more refined items' embeddings based upon the multimodal item-item similarities may positively impact the accuracy performance. Conversely, VBPR's outstanding performance with respect to the other multimodal approaches comes as quite a surprise, considering that more complex and recent models leveraging graph neural networks (such as MMGCN and GRCN) do not outperform it. 

Considering the performance on a dataset level, the most significant variation in metrics between LATTICE and VBPR is observed on \textit{Toys} and \textit{Clothing}, while the difference is reduced on \textit{Office}. Notably, \textit{Toys} and \textit{Clothing} store three and four times more interactions than \textit{Office}, respectively, but they are much sparser. This emphasizes LATTICE's ability to recommend more accurate items despite the higher dataset sparsity. Assessing the other models' performance, MMGCN works exceptionally well on \textit{Toys} but shows the lowest performance as the number of interactions and sparsity increase. GRCN, in contrast, excels with highly sparse data, exhibiting an opposite trend to MMGCN.

From a metric-wise analysis, LATTICE outperforms VBPR in correctly predicting relevant items (high Recall) that are more likely to appear at the top of the recommendation lists (nDCG). However, the same trend is not as evident on the Recall, partly due to its normalization w.r.t. the $k$ recommended items, which can lead to a smaller difference between LATTICE and VBPR as $k$ increases.

\noindent\textbf{Diversity.} As far as recommendation diversity (i.e., iCov) is concerned, the worst-performing model is MMGCN, since its iCov is, in any case, negatively out of scale compared to the other models. For instance, when taking into account \textit{Office}, MMGCN's iCov is slightly better than MostPop (whose item diversity is, by construction, the lowest) demonstrating a restricted ability to engage diverse items in the recommendation lists. Unexpectedly, the second-worst model is LATTICE, even if its performance is still more balanced to the other approaches than MMGCN's one. Indeed, we observe that while MMGCN is affected by poor accuracy due to the lack of item diversity, LATTICE can deal with both accuracy and diversity. 

As an opposite (but noteworthy) trend, we underline that VBPR and GRCN are generally capable of recommending a wider portion of items than MMGCN and LATTICE, independently on the selected top-$k$. Overall, their iCov values are quite comparable to the ones of Random, which should provide (by definition) the highest item coverage from the catalog. We intend to further investigate (and justify) this aspect by assessing the effects of popularity bias.

\noindent\textbf{Popularity bias.} In terms of popularity bias (i.e., APLT), the worst and second-worst models are once again MMGCN and LATTICE (the former on \textit{Office} and \textit{Clothing}, while the latter on \textit{Toys}). As already discussed in~\Cref{sec:aplt}, it makes sense to conceptually bind iCov and APLT. When assessing MMGCN's performance on \textit{Office}, it becomes clear how the model is recommending only a few items (see again the iCov) while achieving good results in terms of accuracy; this demonstrates how the user-item interactions from \textit{Office} may likely be biased towards popular items, and the phenomenon is even amplified due to the dataset small size. The same does not hold on \textit{Clothing} where MMGCN, usually prone to popularity bias, gets also really low performance in terms of accuracy. Conversely, LATTICE can recommend popular items thus pushing its accuracy performance without amplifying the popularity bias phenomenon as much as MMGCN does. Indeed, even if LATTICE's iCov is the second-worst across all the datasets, the metric is always close to the best models in terms of diversity.

Finally, VBPR and GRCN confirm their ability (already observed on the diversity measure) to tackle also popularity bias in all experimental settings. Particularly, while we recognize that VBPR performance is slightly increased with respect to MFBPR in terms of iCov and APLT (the two approaches are almost similar), GRCN results are quite remarkable. It might be the case that its graph edges pruning technique (driven by multimodal signals) is reducing the influence of noisy user-item interactions (i.e., redundant edges which might involve popular items), thus helping to diversify the recommendations by considering also several long-tail items.

\noindent\textsc{\bfseries Summary.} \textit{\ul{In a standard multimodal setting, LATTICE stands out for its accuracy performance and ability to handle dataset sparsity, but at the detriment of amplifying popularity bias; MMGCN struggles with diversity, exhibits strong popularity bias, and sacrifices accuracy in certain scenarios; VBPR and GRCN, in different manners, better manage all the metrics by finding the right compromise among them.}}

\subsection{Modalities influence on recommendation performance (RQ2)}\label{sec:RQ2}
\begin{figure*}[!t]
\centering
\captionsetup[subfigure]{font=footnotesize,labelfont=footnotesize}

\subfloat[Recall]{
\begin{tikzpicture}
\begin{scope}[scale=0.60, transform shape]

\begin{axis}[
legend cell align={left},
legend style={fill opacity=0.8, draw opacity=1, text opacity=1, draw=lightgray204},
tick align=outside,
tick pos=left,
x grid style={darkgray176},
xmin=-0.3, xmax=3.55,
xtick style={color=black},
xtick={0.123,1.123,2.123,3.123},
xticklabels={VBPR,MMGCN,GRCN,LATTICE},
y grid style={darkgray176},
ymin=-30, ymax=30,
ytick style={color=black},
ytick={-20, -10, 0, 10, 20},
yticklabels={-20\%, -10\%, 0\%, +10\%, +20\%},
xlabel={\large \textit{Office}},
]
\draw[draw=none,fill=steelblue31119180] (axis cs:-0.125,0) rectangle (axis cs:0.125,-2.23313916237592);

\draw[draw=none,fill=steelblue31119180] (axis cs:0.875,0) rectangle (axis cs:1.125,0.842683738600947);
\draw[draw=none,fill=steelblue31119180] (axis cs:1.875,0) rectangle (axis cs:2.125,12.1670184588625);
\draw[draw=none,fill=steelblue31119180] (axis cs:2.875,0) rectangle (axis cs:3.125,-1.91747366093645);
\draw[draw=none,fill=darkorange25512714] (axis cs:0.125,0) rectangle (axis cs:0.375,-2.7761333419012);

\draw[draw=none,fill=darkorange25512714] (axis cs:1.125,0) rectangle (axis cs:1.375,-44.4374667655878);
\draw[draw=none,fill=darkorange25512714] (axis cs:2.125,0) rectangle (axis cs:2.375,13.0737534803369);
\draw[draw=none,fill=darkorange25512714] (axis cs:3.125,0) rectangle (axis cs:3.375,-3.51339601683583);

\draw (axis cs:-0.5,0) -- (axis cs:5,0);

\end{axis}

\end{scope}
\end{tikzpicture} \quad
\begin{tikzpicture}

\begin{scope}[scale=0.60, transform shape]

\begin{axis}[
legend cell align={left},
legend style={fill opacity=0.8, draw opacity=1, text opacity=1, draw=lightgray204},
tick align=outside,
tick pos=left,
x grid style={darkgray176},
xmin=-0.3, xmax=3.55,
xtick style={color=black},
xtick={0.123,1.123,2.123,3.123},
xticklabels={VBPR,MMGCN,GRCN,LATTICE},
y grid style={darkgray176},
ymin=-30, ymax=30,
ytick style={color=black},
ytick={-20, -10, 0, 10, 20},
yticklabels={-20\%, -10\%, 0\%, +10\%, +20\%},
xlabel={\large \textit{Toys}},
]
\draw[draw=none,fill=steelblue31119180] (axis cs:-0.125,0) rectangle (axis cs:0.125,-0.234024574174344);

\draw[draw=none,fill=steelblue31119180] (axis cs:0.875,0) rectangle (axis cs:1.125,22.6247278631341);
\draw[draw=none,fill=steelblue31119180] (axis cs:1.875,0) rectangle (axis cs:2.125,-0.717339183913258);
\draw[draw=none,fill=steelblue31119180] (axis cs:2.875,0) rectangle (axis cs:3.125,-8.47146115545028);
\draw[draw=none,fill=darkorange25512714] (axis cs:0.125,0) rectangle (axis cs:0.375,2.77314279725133);

\draw[draw=none,fill=darkorange25512714] (axis cs:1.125,0) rectangle (axis cs:1.375,35.1969838463653);
\draw[draw=none,fill=darkorange25512714] (axis cs:2.125,0) rectangle (axis cs:2.375,10.5537045469473);
\draw[draw=none,fill=darkorange25512714] (axis cs:3.125,0) rectangle (axis cs:3.375,3.09481776023194);

\draw (axis cs:-0.5,0) -- (axis cs:5,0);

\end{axis}

\end{scope}

\end{tikzpicture} \quad
\begin{tikzpicture}

\begin{scope}[scale=0.60, transform shape]

\begin{axis}[
legend cell align={left},
legend style={fill opacity=0.8, draw opacity=1, text opacity=1, draw=lightgray204},
tick align=outside,
tick pos=left,
x grid style={darkgray176},
xmin=-0.3, xmax=3.55,
xtick style={color=black},
xtick={0.123,1.123,2.123,3.123},
xticklabels={VBPR,MMGCN,GRCN,LATTICE},
y grid style={darkgray176},
ymin=-30, ymax=30,,
ytick style={color=black},
ytick={-20, -10, 0, 10, 20},
yticklabels={-20\%, -10\%, 0\%, +10\%, +20\%},
xlabel={\large \textit{Clothing}},
]
\draw[draw=none,fill=steelblue31119180] (axis cs:-0.125,0) rectangle (axis cs:0.125,-0.577398532327521);

\draw[draw=none,fill=steelblue31119180] (axis cs:0.875,0) rectangle (axis cs:1.125,-1.87746693199128);
\draw[draw=none,fill=steelblue31119180] (axis cs:1.875,0) rectangle (axis cs:2.125,-2.65072952638028);
\draw[draw=none,fill=steelblue31119180] (axis cs:2.875,0) rectangle (axis cs:3.125,-13.3025744857898);
\draw[draw=none,fill=darkorange25512714] (axis cs:0.125,0) rectangle (axis cs:0.375,-0.222014841590676);

\draw[draw=none,fill=darkorange25512714] (axis cs:1.125,0) rectangle (axis cs:1.375,6.21804986538689);
\draw[draw=none,fill=darkorange25512714] (axis cs:2.125,0) rectangle (axis cs:2.375,4.64683075788471);
\draw[draw=none,fill=darkorange25512714] (axis cs:3.125,0) rectangle (axis cs:3.375,4.08791383789389);

\draw (axis cs:-0.5,0) -- (axis cs:5,0);

\end{axis}

\end{scope}
\end{tikzpicture}\label{fig:Recall}}

\vspace{0.5cm}

\subfloat[iCov]{
\begin{tikzpicture}

\begin{scope}[scale=0.60, transform shape]

\begin{axis}[
legend cell align={left},
legend style={fill opacity=0.8, draw opacity=1, text opacity=1, draw=lightgray204},
tick align=outside,
tick pos=left,
x grid style={darkgray176},
xmin=-0.3, xmax=3.55,
xtick style={color=black},
xtick={0.123,1.123,2.123,3.123},
xticklabels={VBPR,MMGCN,GRCN,LATTICE},
y grid style={darkgray176},
ymin=-30, ymax=30,,
ytick style={color=black},
ytick={-20, -10, 0, 10, 20},
yticklabels={-20\%, -10\%, 0\%, +10\%, +20\%},
xlabel={\large \textit{Office}},
]
\draw[draw=none,fill=steelblue31119180] (axis cs:-0.125,0) rectangle (axis cs:0.125,-0.0849617672047579);

\draw[draw=none,fill=steelblue31119180] (axis cs:0.875,0) rectangle (axis cs:1.125,3.57142857142857);
\draw[draw=none,fill=steelblue31119180] (axis cs:1.875,0) rectangle (axis cs:2.125,-0.24906600249066);
\draw[draw=none,fill=steelblue31119180] (axis cs:2.875,0) rectangle (axis cs:3.125,-0.259179265658747);
\draw[draw=none,fill=darkorange25512714] (axis cs:0.125,0) rectangle (axis cs:0.375,-0.0424808836023789);

\draw[draw=none,fill=darkorange25512714] (axis cs:1.125,0) rectangle (axis cs:1.375,-65.1785714285714);
\draw[draw=none,fill=darkorange25512714] (axis cs:2.125,0) rectangle (axis cs:2.375,-0.70568700705687);
\draw[draw=none,fill=darkorange25512714] (axis cs:3.125,0) rectangle (axis cs:3.375,-0.561555075593952);

\draw (axis cs:-0.5,0) -- (axis cs:5,0);

\end{axis}

\end{scope}

\end{tikzpicture} \quad
\begin{tikzpicture}

\begin{scope}[scale=0.60, transform shape]

\begin{axis}[
legend cell align={left},
legend style={fill opacity=0.8, draw opacity=1, text opacity=1, draw=lightgray204},
tick align=outside,
tick pos=left,
x grid style={darkgray176},
xmin=-0.3, xmax=3.55,
xtick style={color=black},
xtick={0.123,1.123,2.123,3.123},
xticklabels={VBPR,MMGCN,GRCN,LATTICE},
y grid style={darkgray176},
ymin=-30, ymax=30,,
ytick style={color=black},
ytick={-20, -10, 0, 10, 20},
yticklabels={-20\%, -10\%, 0\%, +10\%, +20\%},
xlabel={\large \textit{Toys}},
]
\draw[draw=none,fill=steelblue31119180] (axis cs:-0.125,0) rectangle (axis cs:0.125,-0.526029385089788);

\draw[draw=none,fill=steelblue31119180] (axis cs:0.875,0) rectangle (axis cs:1.125,-51.4908624559154);
\draw[draw=none,fill=steelblue31119180] (axis cs:1.875,0) rectangle (axis cs:2.125,0.327332242225859);
\draw[draw=none,fill=steelblue31119180] (axis cs:2.875,0) rectangle (axis cs:3.125,2.49878463782207);
\draw[draw=none,fill=darkorange25512714] (axis cs:0.125,0) rectangle (axis cs:0.375,-0.344639941955378);

\draw[draw=none,fill=darkorange25512714] (axis cs:1.125,0) rectangle (axis cs:1.375,-38.5700545046489);
\draw[draw=none,fill=darkorange25512714] (axis cs:2.125,0) rectangle (axis cs:2.375,-4.02274097682832);
\draw[draw=none,fill=darkorange25512714] (axis cs:3.125,0) rectangle (axis cs:3.375,-0.884783665532329);

\draw (axis cs:-0.5,0) -- (axis cs:5,0);

\end{axis}

\end{scope}

\end{tikzpicture} \quad
\begin{tikzpicture}

\begin{scope}[scale=0.60, transform shape]

\begin{axis}[
legend cell align={left},
legend style={fill opacity=0.8, draw opacity=1, text opacity=1, draw=lightgray204},
tick align=outside,
tick pos=left,
x grid style={darkgray176},
xmin=-0.3, xmax=3.55,
xtick style={color=black},
xtick={0.123,1.123,2.123,3.123},
xticklabels={VBPR,MMGCN,GRCN,LATTICE},
y grid style={darkgray176},
ymin=-30, ymax=30,,
ytick style={color=black},
ytick={-20, -10, 0, 10, 20},
yticklabels={-20\%, -10\%, 0\%, +10\%, +20\%},
xlabel={\large \textit{Clothing}},
]
\draw[draw=none,fill=steelblue31119180] (axis cs:-0.125,0) rectangle (axis cs:0.125,-1.45404922121312);

\draw[draw=none,fill=steelblue31119180] (axis cs:0.875,0) rectangle (axis cs:1.125,-20.9148044692737);
\draw[draw=none,fill=steelblue31119180] (axis cs:1.875,0) rectangle (axis cs:2.125,-1.90641247833622);
\draw[draw=none,fill=steelblue31119180] (axis cs:2.875,0) rectangle (axis cs:3.125,-17.0429923594481);
\draw[draw=none,fill=darkorange25512714] (axis cs:0.125,0) rectangle (axis cs:0.375,0.404686838266435);

\draw[draw=none,fill=darkorange25512714] (axis cs:1.125,0) rectangle (axis cs:1.375,-4.43435754189944);
\draw[draw=none,fill=darkorange25512714] (axis cs:2.125,0) rectangle (axis cs:2.375,-2.61298493534195);
\draw[draw=none,fill=darkorange25512714] (axis cs:3.125,0) rectangle (axis cs:3.375,-3.2329798152583);

\draw (axis cs:-0.5,0) -- (axis cs:5,0);

\end{axis}

\end{scope}

\end{tikzpicture}\label{fig:icov}}

\vspace{0.5cm}

\subfloat[APLT]{
\begin{tikzpicture}

\begin{scope}[scale=0.60, transform shape]

\begin{axis}[
legend cell align={left},
legend style={fill opacity=0.8, draw opacity=1, text opacity=1, draw=lightgray204},
tick align=outside,
tick pos=left,
x grid style={darkgray176},
xmin=-0.3, xmax=3.55,
xtick style={color=black},
xtick={0.123,1.123,2.123,3.123},
xticklabels={VBPR,MMGCN,GRCN,LATTICE},
y grid style={darkgray176},
ymin=-30, ymax=30,
ytick style={color=black},
ytick={-20, -10, 0, 10, 20},
yticklabels={-20\%, -10\%, 0\%, +10\%, +20\%},
xlabel={\large \textit{Office}},
]
\draw[draw=none,fill=steelblue31119180] (axis cs:-0.125,0) rectangle (axis cs:0.125,-0.59222384344688);

\draw[draw=none,fill=steelblue31119180] (axis cs:0.875,0) rectangle (axis cs:1.125,-100);
\draw[draw=none,fill=steelblue31119180] (axis cs:1.875,0) rectangle (axis cs:2.125,-4.53743975549548);
\draw[draw=none,fill=steelblue31119180] (axis cs:2.875,0) rectangle (axis cs:3.125,-1.41971605678867);
\draw[draw=none,fill=darkorange25512714] (axis cs:0.125,0) rectangle (axis cs:0.375,-5.00386232941376);

\draw[draw=none,fill=darkorange25512714] (axis cs:1.125,0) rectangle (axis cs:1.375,-100);
\draw[draw=none,fill=darkorange25512714] (axis cs:2.125,0) rectangle (axis cs:2.375,-7.21464676149056);
\draw[draw=none,fill=darkorange25512714] (axis cs:3.125,0) rectangle (axis cs:3.375,7.68846230753845);

\draw (axis cs:-0.5,0) -- (axis cs:5,0);

\end{axis}

\end{scope}

\end{tikzpicture} \quad
\begin{tikzpicture}

\begin{scope}[scale=0.60, transform shape]

\begin{axis}[
legend cell align={left},
legend style={fill opacity=0.8, draw opacity=1, text opacity=1, draw=lightgray204},
tick align=outside,
tick pos=left,
x grid style={darkgray176},
xmin=-0.3, xmax=3.55,
xtick style={color=black},
xtick={0.123,1.123,2.123,3.123},
xticklabels={VBPR,MMGCN,GRCN,LATTICE},
y grid style={darkgray176},
ymin=-30, ymax=30,
ytick style={color=black},
ytick={-20, -10, 0, 10, 20},
yticklabels={-20\%, -10\%, 0\%, +10\%, +20\%},
xlabel={\large \textit{Toys}},
]
\draw[draw=none,fill=steelblue31119180] (axis cs:-0.125,0) rectangle (axis cs:0.125,-1.41850162582115);

\draw[draw=none,fill=steelblue31119180] (axis cs:0.875,0) rectangle (axis cs:1.125,-76.1759910447036);
\draw[draw=none,fill=steelblue31119180] (axis cs:1.875,0) rectangle (axis cs:2.125,-7.7002905770029);
\draw[draw=none,fill=steelblue31119180] (axis cs:2.875,0) rectangle (axis cs:3.125,8.06761284493482);
\draw[draw=none,fill=darkorange25512714] (axis cs:0.125,0) rectangle (axis cs:0.375,3.91724679745975);

\draw[draw=none,fill=darkorange25512714] (axis cs:1.125,0) rectangle (axis cs:1.375,-72.0390334120166);
\draw[draw=none,fill=darkorange25512714] (axis cs:2.125,0) rectangle (axis cs:2.375,-15.5064342050644);
\draw[draw=none,fill=darkorange25512714] (axis cs:3.125,0) rectangle (axis cs:3.375,3.66298987313182);

\draw (axis cs:-0.5,0) -- (axis cs:5,0);

\end{axis}

\end{scope}

\end{tikzpicture} \quad
\begin{tikzpicture}

\begin{scope}[scale=0.60, transform shape]

\begin{axis}[
legend cell align={left},
legend style={fill opacity=0.8, draw opacity=1, text opacity=1, draw=lightgray204},
tick align=outside,
tick pos=left,
x grid style={darkgray176},
xmin=-0.3, xmax=3.55,
xtick style={color=black},
xtick={0.123,1.123,2.123,3.123},
xticklabels={VBPR,MMGCN,GRCN,LATTICE},
y grid style={darkgray176},
ymin=-30, ymax=30,
ytick style={color=black},
ytick={-20, -10, 0, 10, 20},
yticklabels={-20\%, -10\%, 0\%, +10\%, +20\%},
xlabel={\large \textit{Clothing}},
]
\draw[draw=none,fill=steelblue31119180] (axis cs:-0.125,0) rectangle (axis cs:0.125,-9.00922138251403);

\draw[draw=none,fill=steelblue31119180] (axis cs:0.875,0) rectangle (axis cs:1.125,-70.2422145328718);
\draw[draw=none,fill=steelblue31119180] (axis cs:1.875,0) rectangle (axis cs:2.125,-20.2344330663143);
\draw[draw=none,fill=steelblue31119180] (axis cs:2.875,0) rectangle (axis cs:3.125,-34.2155241070881);
\draw[draw=none,fill=darkorange25512714] (axis cs:0.125,0) rectangle (axis cs:0.375,3.39180475629209);

\draw[draw=none,fill=darkorange25512714] (axis cs:1.125,0) rectangle (axis cs:1.375,-57.8431372549027);
\draw[draw=none,fill=darkorange25512714] (axis cs:2.125,0) rectangle (axis cs:2.375,-24.7568956974152);
\draw[draw=none,fill=darkorange25512714] (axis cs:3.125,0) rectangle (axis cs:3.375,-0.275684616798511);

\draw (axis cs:-0.5,0) -- (axis cs:5,0);

\end{axis}

\end{scope}

\end{tikzpicture}\label{fig:APLT}}

\vspace{0.5em}

\begin{tikzpicture}
        \begin{customlegend}[legend columns=2,
        legend entries={visual, textual}]
        \addlegendimage{area legend, fill=steelblue31119180}
        \addlegendimage{area legend, fill=darkorange25512714}
        \end{customlegend}
    \end{tikzpicture}

\caption{Percentage variation on the (a) Recall, (b) iCov, and (c) APLT when training the multimodal recommender systems with either \textcolor{steelblue31119180}{visual} or \textcolor{darkorange25512714}{textual} modalities. The 0\% line stands for the reference performance provided by the multimodal version of the model. All results refer to the top@20 recommendation lists.}\label{fig:rq2}
\end{figure*}
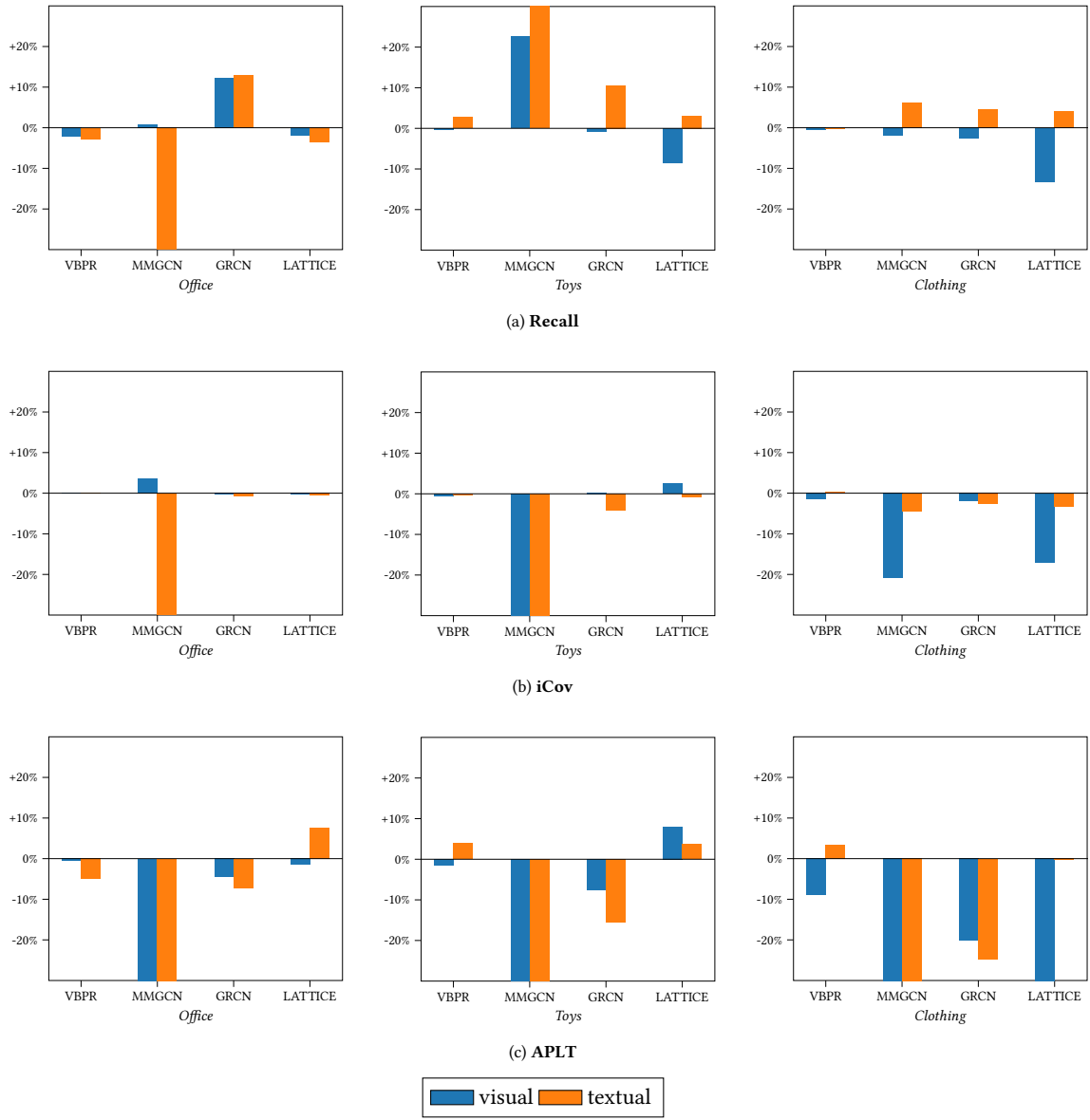
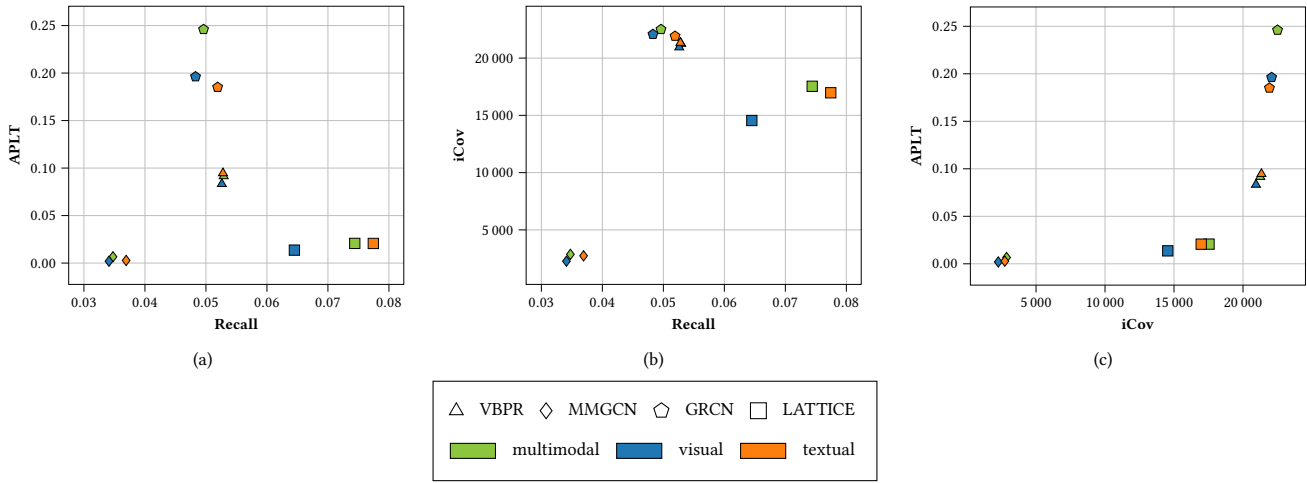
\begin{figure*}[!t]
\centering
\captionsetup[subfigure]{font=footnotesize,labelfont=footnotesize}
\subfloat[]{
\begin{tikzpicture}

\begin{scope}[scale=0.65, transform shape]

\begin{axis}[
yticklabel style={
/pgf/number format/.cd,
fixed, fixed zerofill,
precision=2
},
xmajorgrids,
ymajorgrids,
scaled y ticks=false,
xticklabel style={
/pgf/number format/.cd,
fixed, fixed zerofill,
precision=2
},
scaled x ticks=false,
tick align=outside,
tick pos=left,
xlabel={\textbf{Recall}},
xtick style={color=black},
ylabel={\textbf{APLT}},
ytick style={color=black},
xmin=0.0275
]

\addplot [LimeGreen, draw=black, mark=triangle*, mark size=3pt]
table{%
x  y
0.0529123887681279 0.091546703226953
};
\addplot [steelblue31119180, draw=black, mark=triangle*, mark size=3pt]
table{%
x  y
0.0526068734119613 0.0832990580648437
};

\addplot [darkorange25512714, draw=black, mark=triangle*, mark size=3pt]
table{%
x  y
0.0527949154120225 0.0946517886612334
};

\addplot [LimeGreen, draw=black, mark=diamond*, mark size=3pt]
table{%
x  y
0.0347594514134235 0.0066037017289968
};
\addplot [steelblue31119180, draw=black, mark=diamond*, mark size=3pt]
table{%
x  y
0.0341068542073949 0.0019651153934039
};
\addplot [darkorange25512714, draw=black, mark=diamond*, mark size=3pt]
table{%
x  y
0.0369208114352451 0.0027839134739888
};

\addplot [LimeGreen, draw=black, mark=pentagon*, mark size=3pt]
table{%
x  y
0.0496086825472155 0.2459491710462843
};
\addplot [steelblue31119180, draw=black, mark=pentagon*, mark size=3pt]
table{%
x  y
0.0482936905512882 0.196182750653769
};
\addplot [darkorange25512714, draw=black, mark=pentagon*, mark size=3pt]
table{%
x  y
0.0519139140664009 0.1850597913016985
};

\addplot [LimeGreen, draw=black, mark=square*, mark size=3pt]
table{%
x  y
0.0744010291043576 0.0207213039835478
};
\addplot [steelblue31119180, draw=black, mark=square*, mark size=3pt]
table{%
x  y
0.0645037767895563 0.013631401223754
};
\addplot [darkorange25512714, draw=black, mark=square*, mark size=3pt]
table{%
x  y
0.0774424790686501 0.0206641785360651
};

\end{axis}

\end{scope}

\end{tikzpicture}
    \label{fig:recall_aplt}} \quad
\subfloat[]{
\begin{tikzpicture}

\begin{scope}[scale=0.65, transform shape]

\begin{axis}[
xmajorgrids,
ymajorgrids,
xticklabel style={
/pgf/number format/.cd,
fixed, fixed zerofill,
precision=2
},
scaled x ticks=false,
tick align=outside,
tick pos=left,
xlabel={\textbf{Recall}},
xtick style={color=black},
ylabel={\textbf{iCov}},
ytick style={color=black},
yticklabels={5\,000, 10\,000, 15\,000, 20\,000},
ytick={5000, 10000, 15000, 20000},
scaled y ticks=false,
xmin=0.0275
]

\addplot [LimeGreen, draw=black, mark=triangle*, mark size=3pt]
table{%
x  y
0.0529123887681279 21251
};
\addplot [steelblue31119180, draw=black, mark=triangle*, mark size=3pt]
table{%
x  y
0.0526068734119613 20942
};
\addplot [darkorange25512714, draw=black, mark=triangle*, mark size=3pt]
table{%
x  y
0.0527949154120225 21337
};

\addplot [LimeGreen, draw=black, mark=diamond*, mark size=3pt]
table{%
x  y
0.0347594514134235 2864
};
\addplot [steelblue31119180, draw=black, mark=diamond*, mark size=3pt]
table{%
x  y
0.0341068542073949 2265
};
\addplot [darkorange25512714, draw=black, mark=diamond*, mark size=3pt]
table{%
x  y
0.0369208114352451 2737
};


\addplot [LimeGreen, draw=black, mark=pentagon*, mark size=3pt]
table{%
x  y
0.0496086825472155 22503
};
\addplot [steelblue31119180, draw=black, mark=pentagon*, mark size=3pt]
table{%
x  y
0.0482936905512882 22074
};
\addplot [darkorange25512714, draw=black, mark=pentagon*, mark size=3pt]
table{%
x  y
0.0519139140664009 21915
};

\addplot [LimeGreen, draw=black, mark=square*, mark size=3pt]
table{%
x  y
0.0744010291043576 17538
};
\addplot [steelblue31119180, draw=black, mark=square*, mark size=3pt]
table{%
x  y
0.0645037767895563 14549
};
\addplot [darkorange25512714, draw=black, mark=square*, mark size=3pt]
table{%
x  y
0.0774424790686501 16971
};

\end{axis}

\end{scope}

\end{tikzpicture}
    \label{fig:recall_icov}} \quad
\subfloat[]{
\begin{tikzpicture}

\begin{scope}[scale=0.65, transform shape]

\begin{axis}[
xmajorgrids,
ymajorgrids,
yticklabel style={
/pgf/number format/.cd,
fixed, fixed zerofill,
precision=2
},
scaled y ticks=false,
tick align=outside,
tick pos=left,
xlabel={\textbf{iCov}},
xtick style={color=black},
ylabel={\textbf{APLT}},
ytick style={color=black},
xticklabels={5\,000, 10\,000, 15\,000, 20\,000},
xtick={5000, 10000, 15000, 20000},
scaled x ticks=false
]

\addplot [LimeGreen, draw=black, mark=triangle*, mark size=3pt]
table{%
x  y
21251 0.091546703226953
};
\addplot [steelblue31119180, draw=black, mark=triangle*, mark size=3pt]
table{%
x  y
20942 0.0832990580648437
};
\addplot [darkorange25512714, draw=black, mark=triangle*, mark size=3pt]
table{%
x  y
21337 0.0946517886612334
};

\addplot [LimeGreen, draw=black, mark=diamond*, mark size=3pt]
table{%
x  y
2864 0.0066037017289968
};
\addplot [steelblue31119180, draw=black, mark=diamond*, mark size=3pt]
table{%
x  y
2265 0.0019651153934039
};
\addplot [darkorange25512714, draw=black, mark=diamond*, mark size=3pt]
table{%
x  y
2737 0.0027839134739888
};

\addplot [LimeGreen, draw=black, mark=pentagon*, mark size=3pt]
table{%
x  y
22503 0.2459491710462843
};
\addplot [steelblue31119180, draw=black, mark=pentagon*, mark size=3pt]
table{%
x  y
22074 0.196182750653769
};
\addplot [darkorange25512714, draw=black, mark=pentagon*, mark size=3pt]
table{%
x  y
21915 0.1850597913016985
};

\addplot [LimeGreen, draw=black, mark=square*, mark size=3pt]
table{%
x  y
17538 0.0207213039835478
};
\addplot [steelblue31119180, draw=black, mark=square*, mark size=3pt]
table{%
x  y
14549 0.013631401223754
};
\addplot [darkorange25512714, draw=black, mark=square*, mark size=3pt]
table{%
x  y
16971 0.0206641785360651
};

\end{axis}

\end{scope}

\end{tikzpicture}
    \label{fig:icov_aplt}}

\vspace{0.3em}

\begin{tikzpicture}
        \begin{customlegend}[legend columns=4,
        legend style={
        at={(0, 0.2)},
       anchor=north,
       inner sep=3pt,
       draw=none,
       style={column sep=0.15cm},
       name=leg1},
        legend entries={\footnotesize VBPR, \footnotesize MMGCN, \footnotesize GRCN, \footnotesize LATTICE}]
        \addlegendimage{only marks, white, draw=black, mark=triangle*, mark size=3pt}
        \addlegendimage{only marks, white, draw=black, mark=diamond*, mark size=3pt}
        \addlegendimage{only marks, white, draw=black, mark=pentagon*, mark size=3pt}
        \addlegendimage{only marks, white, draw=black, mark=square*, mark size=3pt}
        \end{customlegend}
        \begin{customlegend}[legend columns=3,
        legend style={
        at={(0, -0.33)},
       anchor=north,
       inner sep=3pt,
       draw=none,
       style={column sep=0.15cm},
       name=leg2},
        legend entries={\footnotesize multimodal, \footnotesize visual, \footnotesize textual}]
        \addlegendimage{area legend, fill=LimeGreen}
        \addlegendimage{area legend, fill=steelblue31119180}
        \addlegendimage{area legend, fill=darkorange25512714}
        \end{customlegend}
        \node [fit=(leg1)(leg2),draw] {};
    \end{tikzpicture}

\caption{Performance analysis on \textit{Clothing} when comparing (a) Recall \textit{vs.} APLT, (b) Recall \textit{vs.} iCov, and (c) iCov \textit{vs.} APLT for different modality settings involving the \textcolor{LimeGreen}{multimodal}, \textcolor{steelblue31119180}{visual}, and \textcolor{darkorange25512714}{textual} modalities. Metrics are on top@20.}\label{fig:rq2_2}
\end{figure*}
While the previous section has answered how multimodal recommender systems perform in terms of accuracy, diversity, and popularity bias when leveraging the \textit{\textbf{full}} modalities, in the following, we discuss the influence of each \textbf{\textit{single}} modality on the performance. We consider two evaluation dimensions where modalities influence is assessed (i) on accuracy, diversity, and popularity bias separately, and (ii) on pairs of metrics to investigate their joint variations. 

\noindent\textbf{Modalities influence on the single metric.} 
\Cref{fig:rq2} displays the influence of each modality calculated as percentage variation with respect to the multimodal baseline, on the top@$20$ recommendation lists. We select the Recall (\Cref{fig:Recall}), iCov (\Cref{fig:icov}), and APLT (\Cref{fig:APLT}) for accuracy, diversity, and popularity bias, respectively.

As regards the accuracy performance (\Cref{fig:Recall}), we notice how the trend is not consistent across all the datasets and models. Particularly, when considering \textit{Office}, we observe that only VBPR and LATTICE fully exploit multimodality (indeed, their performance decreases when the modalities are injected separately); on an opposite level, on MMGCN, the visual modality slightly improves the multimodal setting, while the textual modality even worsens it; then, GRCN achieves better performance on both the visual and textual modalities, suggesting that this approach may not take advantage of the multimodal configuration. On the \textit{Toys} dataset, the only textual setting generally improves the performance, bringing important information to the model learning interaction. The model benefiting from the single modality the most is MMGCN, which has an improvement of at least 20\% on both visual and textual. For the remaining models, the trend is quite stable with the textual and visual modalities improving and reducing the performance, respectively. Finally, we observe that \textit{Clothing} is the only dataset showing consistent trends. Indeed, the visual modality reduces the Recall while the textual increases it (with the only exception of VBPR whose percentage variation is negligible).

Differently from the accuracy analysis, we recognize a quasi-stable trend in the performance variation measured for the diversity metric (\Cref{fig:icov}). Considering the \textit{Office} dataset, each modality's contribution is generally irrelevant except for MMGCN, for which the visual modality slightly improves the coverage across the whole recommendation list, while the textual one worsens the performance by a large margin. Assessing the trend on \textit{Toys}, both the modalities decrease the coverage performance of the model when injected separately in the recommendation pipeline; remarkably, MMGCN is once again the model affected by the single modality presence the most, but this time the coverage performance widely deteriorates because of both the visual and textual modalities. Finally, on \textit{Clothing}, both modalities lower the model's item coverage, with specific reference to the visual modality.

As the last part of our analysis, we take into account each modality's contribution to the popularity bias dimension (\Cref{fig:APLT}). Starting from \textit{Office}, we notice how both modalities are prone to enforce popularity bias if injected singularly, with the only exception of LATTICE whose textual modality limits the popularity bias (the APLT increases); this is interesting as we remind that LATTICE is the second-worst model in terms of popularity bias, but using only the textual modality reduces its accuracy performance and the influence of popular items in the recommendation list. When it comes to the \textit{Toys} dataset, every single modality enforces the popularity bias of MMGCN and GRCN; for VBPR, the visual and textual modalities amplify and reduce the bias, respectively, while for LATTICE both the visual and textual modalities limit the popularity bias. Finally, on \textit{Clothing}, both the modalities show to increase the popularity bias of the model (but the textual one on VBPR and LATTICE).

\noindent\textbf{Modalities cross-influence on metrics pairs.}
To conclude, we discuss the cross-influence of each modality setting (i.e., visual, textual, and multimodal) on pairs of metrics. In this respect, we decide to display (\Cref{fig:rq2_2}) the joint trend of (a) accuracy and popularity bias (i.e., Recall \textit{vs.} APLT), (b) accuracy and diversity (i.e., Recall \textit{vs.} iCov), and (c) diversity and popularity bias (i.e., iCov \textit{vs.} APLT). We only report the results on \textit{Clothing} for top@20 recommendations.

In detail, VBPR and MMGCN are the models being affected by each specific modality the least, since the performance measures assessed on visual and textual are generally aligned with the multimodal reference. Regarding LATTICE, we notice that the textual modality has a major accuracy influence with respect to popularity bias and diversity. Indeed, the textual modality improves the Recall without having a relevant effect in terms of iCov and APLT; conversely, the visual modality reduces the accuracy by jointly worsening the diversity and the popularity bias. Finally, when considering GRCN, we observe that the multimodal setting reduces the popularity bias without affecting the accuracy and diversity.

\noindent\textsc{\bfseries Summary.} \textit{\ul{In a single modality setting, the textual one improves the accuracy, while both modalities negatively affect the diversity and reinforce the popularity bias. When evaluating the modalities' influence across metrics pairs, the textual modality has a significant influence on accuracy but minimal effects on diversity and popularity bias; conversely, the visual modality reduces accuracy and jointly worsens the popularity bias and diversity.}}

\section{Conclusion and Future Work}
Motivated by the assumption that factorization models in recommendation (such as MFBPR) are affected by popularity bias, in this work, we provided one of the first systematic analyses on how multimodal-aware recommender systems (largely built upon MFBPR) further amplify the recommendation of popular items. After having selected four state-of-the-art multimodal recommender systems, namely, VBPR, MMGCN, GRCN, and LATTICE, we proposed an exhaustive experimental study involving three datasets from the Amazon catalog, four metrics spanning three evaluation dimensions (i.e., accuracy, diversity, and popularity bias), and three modalities settings (i.e., multimodal, only visual, and only textual). Results demonstrated that, in a standard multimodal setting, VBPR and GRCN can strike a better compromise between all evaluated metrics than MMGCN and LATTICE; furthermore, the separate injection of the visual and textual modalities can improve the accuracy but negatively impact the diversity and popularity bias. Conclusively, a complementary investigation regarding the modalities' influence on metrics pairs outlined that the textual modality has a considerable impact on accuracy but little effect on diversity and popularity bias, whereas the visual modality reduces accuracy while exacerbating popularity bias and limiting the diversity. Such findings pave the way to a more complete study on the performance of these models and other more recent approaches in multimodal recommendation.

\begin{acks}
This work was partially supported by the following projects: Secure Safe Apulia,
MISE CUP: I14E20000020001 CTEMT - Casa delle Tecnologie Emergenti Comune di Matera, CT\_FINCONS\_III, OVS Fashion Retail Reloaded, LUTECH DIGITALE 4.0, KOINÈ.
\end{acks}

\bibliographystyle{ACM-Reference-Format}
\bibliography{main}

\end{document}